\documentclass[12pt, a4paper]{amsart}

\usepackage[utf8]{inputenc}
\usepackage[T1]{fontenc}
\usepackage{lmodern}
\usepackage{amsmath,amssymb} 
\usepackage{graphicx}
\usepackage{booktabs}
\usepackage{xurl}            
\usepackage[margin=1in]{geometry} 
\usepackage{amsthm}
\usepackage{physics}
\usepackage{braket}
\usepackage{comment}
\usepackage{microtype} 
     
\usepackage{hyperref}       

\usepackage[
  backend=biber,             
  citestyle=authoryear,      
  bibstyle=numeric,          
  maxcitenames=2,            
  maxbibnames=99             
]{biblatex}

\usepackage[dvipsnames]{xcolor}
\usepackage{enumitem}

\setlist[enumerate]{
  before=\vspace{0.5ex},
  after=\vspace{0.5ex},
  topsep=0.1ex,
  itemsep=0.5ex,
  parsep=0ex,
}

\setlist[itemize]{
  before=\vspace{0.5ex},
  after=\vspace{0.5ex},
  topsep=0.1ex,
  itemsep=1ex,
  parsep=0ex,
}

\makeatletter
\renewcommand{\subsection}{\@startsection{subsection}{2}{\z@}%
  {-3.25ex\@plus -1ex \@minus -.2ex}%
  {1.5ex \@plus .2ex}%
  {\normalfont\normalsize\bfseries}} 
\makeatother

\makeatletter
\renewcommand{\subsubsection}{\@startsection{subsubsection}{3}{\z@}%
  {-3.25ex\@plus -1ex \@minus -.2ex}%
  {1.5ex \@plus .2ex}%
  {\normalfont\normalsize\bfseries}} 
\makeatother

\makeatletter
\renewcommand{\paragraph}{\@startsection{paragraph}{4}{\z@}%
  {3.25ex \@plus 1ex \@minus .2ex}%
  {0.5ex}
  {\normalfont\normalsize\bfseries}}
\makeatother

\theoremstyle{plain}

\theoremstyle{remark}

\theoremstyle{definition}
\newtheorem{definition}{Definition}

\newcommand{\MPG}{\mathrm{MPG}}

\newcommand{\R}{\mathbb{R}}
\newcommand{\C}{\mathbb{C}}
\newcommand{\N}{\mathbb{N}}

\definecolor{light-gray}{gray}{0.95}

\selectfont

\title[Rogue Variable Theory: A Quantum-Compatible Cognition Framework]%
{Rogue Variable Theory: A Quantum-Compatible Cognition Framework with a Rosetta Stone Alignment Algorithm}

\author[J. Małecki]{Jacek Małecki}
\address{Department of Mathematics, Faculty of Mathematics, Wrocław University of Science and Technology}
\email{jacek.malecki@pwr.edu.pl}

\author[A. Mathiesen-Ohman]{Alexander Mathiesen-Ohman}
\address{AMOTHO Research Institute, Vallsjön 20, 780 00 Rörbäcksnäs, Sweden}
\email{amo@amotho.com}

\dedicatory{\today}


\begin{document}

\begin{abstract}
    Many of the most consequential dynamics in human cognition occur \emph{before} events become explicit: before decisions are finalized, emotions are labeled, or meanings stabilize into narrative form. These pre-event states are characterized by ambiguity, contextual tension, and competing latent interpretations. Rogue Variable Theory (RVT) formalizes such states as \emph{Rogue Variables}: structured, pre-event cognitive configurations that influence outcomes while remaining unresolved or incompatible with a system's current representational manifold. We present a quantum-consistent information-theoretic implementation of RVT based on a time-indexed \emph{Mirrored Personal Graph} (MPG) embedded into a fixed graph Hilbert space, a normalized \emph{Quantum MPG State} (QMS) constructed from node and edge metrics under context, Hamiltonian dynamics derived from graph couplings, and an error-weighted `rogue operator'' whose principal eigenvectors identify rogue factor directions and candidate Rogue Variable segments. We further introduce a \emph{Rosetta Stone Layer} (RSL) that maps user-specific latent factor coordinates into a shared reference Hilbert space to enable cross-user comparison and aggregation without explicit node alignment. The framework is fully implementable on classical systems and does not assume physical quantum processes; \emph{collapse} is interpreted as informational decoherence under interaction, often human clarification.
\end{abstract}

\footnotetext[1]{{Alexander Mathiesen-Ohman, \textit{Systems and Methods for Quantum-Inspired Rogue Variable Modelling (QRVM), Human-in-the-Loop Decoherence, and Collective Cognitive Inference in Human-AI Symbiotic Architectures}, Provisional Utility Patent No 63/941,301}}

\maketitle


\section{Introduction: The problem of pre-event states}

    \subsection{The blind spot in modern AI and cognitive science}
    Most AI systems and cognitive models are optimized around \emph{post-event artifacts}: expressed decisions, labeled emotions, observable behavior, and articulated beliefs. Yet a large fraction of human experience unfolds in a qualitatively different regime: \emph{pre-event states} that arise \emph{before} commitments crystallize into discrete reports, actions, or stable narratives. Typical examples include hesitation before choice, tension before conflict, intuition before articulation, creative uncertainty before insight, and ethical discomfort before moral reasoning.

    \medskip

    These pre-event states are neither noise nor error. They are often \emph{structurally meaningful} and predictive of imminent transitions, but they are transient, context-dependent, and difficult to represent with conventional state variables. In practice, they frequently appear as ``incoherence'' or ``instability'' relative to a model that expects a single settled interpretation.

    \medskip

     A core difficulty is that many classical modeling assumptions are systematically violated in human cognition. Standard probabilistic and dynamical approaches commonly presume (i) mutually exclusive hypotheses, as if a cognitive state always corresponds to one determinate interpretation; (ii) additive probabilities defined over a fixed event space, as if the meaning of propositions is stable across contexts; (iii) commutative inference, such that the final belief state depends only on the set of information received and not on the order in which it is processed; and (iv) stable, approximately Markovian state transitions, where the next cognitive state depends only on the current state rather than on the trajectory by which it was reached. Empirical cognition departs from these assumptions in systematic ways: people often occupy partially committed or ambiguous states prior to a settled report; contextual framing changes what features are attended to and even what counts as relevant evidence; sequential evaluation produces order effects (e.g., primacy/recency and question-order dependence), indicating non-commutative updating; and competing interpretations can interact in ways that are not well captured by simple mixture models or memoryless dynamics. As a result, classical updating rules and low-order dynamical models can mischaracterize cognition precisely in regimes dominated by context sensitivity, path dependence, and instability before commitment ( \cite{BusemeyerBruza2012,PothosBusemeyer2013,Khrennikov2010,HogarthEinhorn1992,TverskyKahneman1981}).

    \subsection{Rogue Variables as pre-event cognitive structures}
    Rogue Variable Theory (RVT) addresses this blind spot by treating pre-event cognition as a first-class object of representation. Informally, a \emph{Rogue Variable} is a structured cognitive configuration that influences downstream behavior while remaining unresolved or incompatible with the model's current representational manifold. The key point is that unresolvedness can be informative: what has not yet been decided, named, or stabilized can nevertheless shape what comes next.

    \medskip
    
    To operationalize RVT, we require a representation that can encode (i) simultaneous competing tendencies, (ii) context-sensitive updates, and (iii) transitions from ambiguity to commitment under interaction. In this paper we instantiate RVT using a quantum-consistent information-theoretic formalism that is fully implementable on classical hardware and makes no claims about physical quantum substrates. Within this stance, ``collapse'' refers to \emph{informational decoherence}: a clarification-driven update that reduces ambiguity by interaction (often human-in-the-loop).

    \subsection{From conceptual theory to implementable mathematics}
    Our technical approach connects the conceptual RVT picture to a concrete architecture built on three components.

    First, we represent an evolving personal cognitive structure as a time-indexed \emph{Mirrored Personal Graph} (MPG). At each time $t$, the MPG snapshot $G_t=(V_t,E_t)$ contains nodes and directed edges augmented with interpretable metrics. We embed all time-varying snapshots into a fixed graph Hilbert space $\ell^2(V)$ over a global node universe, and construct a normalized \emph{Quantum MPG State} (QMS) $\ket{\Psi_t}$ whose amplitudes are obtained from node/edge metrics under context.

    \medskip
    
    Second, we introduce Hamiltonian dynamics on the graph Hilbert space, where a self-adjoint operator $\hat{H}_t$ constructed from edge couplings and node-local terms governs continuous-time evolution between observation-driven reconstructions. A baseline Hamiltonian $\hat{H}_{\mathrm{base}}$ provides a spectral latent-factor basis, enabling interpretable tracking of dynamical modes.

    \medskip
    
    Third, we define an error-weighted operator $\hat{O}_{\epsilon}$ built from QMS snapshots weighted by a divergence signal $\epsilon_t$. The principal eigenvectors of $\hat{O}_{\epsilon}$ identify \emph{rogue factor directions} associated with high error, and their high-loading node sets define candidate Rogue Variable segments. These candidates can be evaluated by ablation: removing a segment's contribution from $\ket{\Psi_t}$ and measuring average error reduction.

    \subsection{Cross-user generalization via the Rosetta Stone Layer}

    A central goal of RVT is to discover recurring pre-event configurations and transition signatures across individuals without requiring brittle node-level alignment between personal graphs. To this end, we introduce a \emph{Rosetta Stone Layer} (RSL) alignment algorithm that maps user-specific latent-factor coordinates into a shared reference Hilbert space. User-specific alignment operators enable comparison and aggregation of trajectories, rogue directions, and operators at a common meta-level, while still allowing projections back to each user's native graph space for interpretability and personalization.

    \subsection{Contributions and paper organization}

    The paper makes the following contributions:
        \begin{enumerate}[label=(\roman*)]
            \item A formal RVT instantiation in which time-indexed Mirrored Personal Graphs are embedded into a fixed graph Hilbert space, and pre-event cognition is represented by a normalized Quantum MPG State.
            \item A graph-derived Hamiltonian dynamics framework, including a baseline spectral decomposition that yields interpretable latent dynamical factors.
            \item An operator-theoretic Rogue Variable detection mechanism based on an error-weighted mixture operator whose eigenstructure reveals rogue directions and candidate Rogue Variable segments, together with an ablation-based validation criterion.
            \item A Rosetta Stone Layer alignment algorithm that maps user-specific latent structures into a shared reference space for cross-user comparison, aggregation, and projection back to personalized representations.
        \end{enumerate}

    The remainder of the paper is organized as follows. Section~\ref{sec:RV:intro} introduces Rogue Variables as structured pre-event cognitive states and clarifies how they differ from noise and anomalies. Section~\ref{sec:why_fail} summarizes why standard classical probabilistic and dynamical assumptions can fail in the pre-event regime. Sections~\ref{sec:decoherence}--\ref{sec:RVT_vs_QP} refine the interpretational stance: collapse is treated operationally as informational decoherence under interaction (often human-in-the-loop), and we state the scope of the framework and its non-claims with respect to physical quantum processes. Sections~\ref{sec:math:Hilbert} and~\ref{sec:math:Hamiltonian} then develop the implementable mathematical core: the Quantum MPG representation in a fixed graph Hilbert space and the associated Hamiltonian dynamics. Section~\ref{sec:math:RV} provides an operator-theoretic definition of Rogue Variable directions and segments, together with an ablation-based validation criterion. Section~\ref{sec:math:Rosetta} introduces the Rosetta Stone Layer alignment algorithm for cross-user mapping into a shared reference space. Finally, Sections~\ref{sec:applications} and~\ref{sec:RVT_vs_science} discuss scientific and practical implications and position RVT in the broader scientific landscape, before Section~\ref{sec:conclusion} concludes.

    \subsection{Scope and limitations}
    This manuscript presents a formal modeling proposal and an operator-theoretic detection criterion. It does not provide empirical validation on behavioral or physiological datasets. The mapping $\Psi_{\mathrm{map}}$ from node metrics to amplitudes, the contextual configuration $\mathcal{C}_t$, and the divergence/error signal $\epsilon_t$ are application-defined and must be specified for any concrete deployment. Accordingly, the claims of this paper concern representational adequacy for contextual and order-sensitive inference and the computability of rogue directions and ablation criteria within the proposed Hilbert-space architecture, rather than confirmed psychological mechanisms or performance guarantees in specific tasks.


\section{Rogue Variables and pre-event cognition}
\label{sec:RV:intro}

    \subsection{Conceptual definition}
    We use the term \emph{Rogue Variable} to denote a specific class of pre-event cognitive states that are both influential and not yet settled.

    \begin{definition}[Rogue Variable (conceptual)]
        A \emph{Rogue Variable} is a structured, pre-event cognitive state that exerts influence on future outcomes while remaining unresolved, ambiguous, or incompatible with a system's current representational manifold.
    \end{definition}

    This definition is intended to capture ambiguity as a \emph{signal} rather than a defect. Rogue Variables are not post-hoc labels assigned to outcomes; they are the configurations that precede and shape outcomes. They often appear in human experience as hesitation before choice, tension before conflict, intuition before articulation, creative uncertainty before insight, or ethical discomfort before moral reasoning.

    \medskip

    The following properties are characteristic:
        \begin{itemize}
            \item \textbf{Pre-event.} Rogue Variables arise before explicit decisions, behaviors, or stable narratives, and can therefore provide early information about impending transitions.
            \item \textbf{Context-dependent.} Their interpretation depends on interaction, framing, and sequencing of evidence. What a Rogue Variable ``is'' becomes clearer only under contextual probing, often involving human clarification.
            \item \textbf{High informational value.} Rogue Variables are frequently associated with imminent reorganization of beliefs, commitments, identity fragments, or behavioral trajectories; they are therefore candidates for monitoring and intervention.
            \item \textbf{Non-classical structure.} Rogue Variables typically cannot be faithfully represented as a single settled hypothesis or as a single classical probability distribution over mutually exclusive alternatives. Competing interpretations may co-exist and interfere, and the order of queries or interactions can matter.
        \end{itemize}

    \subsection{Rogue Variables vs.\ noise and anomalies}
    It is important to distinguish Rogue Variables from phenomena that are commonly treated as modeling artifacts.

        \paragraph{Noise.}
        Noise is typically random, low-information variation that does not carry stable structure. It is often treated as error, averaged out, or suppressed by smoothing.

        \paragraph{Anomalies and outliers.}
        Anomalies are often defined relative to a learned baseline and are commonly detected post-hoc as rare deviations. They may be discarded, flagged for quality control, or treated as failures of fit.

        \paragraph{Rogue Variables.}
        Rogue Variables, in contrast, are \emph{structured ambiguities}: transitional, informative configurations that occur \emph{before} a system has enough evidence (or the right kind of interaction) to collapse the state into a stable interpretation. They function as early warnings and are most valuable precisely when classical models label them as incoherence.

        \medskip    

    \noindent The distinction can be summarized as follows:
        \begin{center}
            \begin{tabular}{@{}lll@{}}
            \toprule
            \textbf{Noise} & \textbf{Anomaly} & \textbf{Rogue Variable} \\
            \midrule
            Random fluctuation & Outlier relative to baseline & Structured ambiguity \\
            Low informational content & Often discarded post-hoc & High predictive value \\
            Typically static variation & Often event-labeled after detection & Transitional / pre-event \\
            Treated as error & Treated as failure or exception & Treated as signal \\
            \bottomrule
            \end{tabular}
        \end{center}

    \subsection{From conceptual definition to operational representation}
    The conceptual definition above does not yet specify how Rogue Variables should be represented or detected in an AI system. Our central modeling choice is to treat the pre-event regime as a \emph{state space of competing tendencies} whose evolution depends on context and interaction. In the next section, we introduce a concrete, quantum-consistent representation in which the user's cognitive configuration is embedded into a graph-based state space and updated over time. This provides a foundation on which Rogue Variables can later be defined and detected operationally as directions and segments associated with disproportionate divergence from a baseline manifold.


\section{Why classical models fail}
\label{sec:why_fail}

    \subsection{Limitations of classical probability}

    Many influential approaches in AI and cognitive science are built on classical probability and classical dynamical assumptions. In their standard forms, Bayesian and Markovian models typically rely on some combination of the following premises:
        \begin{itemize}
            \item \textbf{Exclusive hypotheses.} Competing explanations are treated as mutually exclusive elements of a hypothesis set.
            \item \textbf{Additive normalization.} Probabilities are assigned on a single sample space with a total-probability normalization and additive set structure.
            \item \textbf{Independence or simple correlation structure.} Dependencies are assumed to factorize or to be captured by relatively simple correlation or conditional-independence relations.
            \item \textbf{Order-invariant inference.} Inference is often treated as commutative in practice: the order in which information is queried or processed is not expected to change the final state, except through changes in available evidence.
        \end{itemize}

    However, human cognition routinely exhibits systematic deviations from these assumptions. Empirical and theoretical work in cognitive science documents phenomena such as \cite{TverskyKahneman1981,HogarthEinhorn1992,LichtensteinSlovic1971,BusemeyerBruza2012,PothosBusemeyer2013}:
        \begin{itemize}
            \item \textbf{Order effects,} where responses depend on the sequence of questions or prompts;
            \item \textbf{Context sensitivity,} where meaning and judgment shift under framing and interaction;
            \item \textbf{Interference between interpretations,} where multiple latent readings co-exist and modulate each other rather than simply competing as exclusive alternatives;
            \item \textbf{Preference reversals,} where preferences change under elicitation order or representation;
            \item \textbf{Framing dependence,} where semantically equivalent descriptions yield different judgments.
        \end{itemize}
    These effects are not rare edge cases; they are characteristic of cognition under ambiguity-precisely the regime in which Rogue Variables arise. When a system is forced into a single classical sample space with commutative updating, such effects are often treated as noise or irrationality. In RVT, they are treated as signatures of pre-event structure that demands a representation capable of contextual and order-dependent inference.

    \subsection{Premature collapse in AI systems}

    A second, engineering-driven failure mode concerns how contemporary AI systems behave under partial information. A common operational loop is:
        \begin{enumerate}
            \item observe partial or noisy data,
            \item select a most likely interpretation,
            \item act with high confidence.
        \end{enumerate}
    This pattern is often effective when the underlying state is well determined and the uncertainty is reducible. However, when a Rogue Variable is present, this pipeline tends to \emph{collapse ambiguity too early}. The result can be overconfident outputs, hallucinated certainty, brittle decision-making, and, in high-stakes settings, ethical or value misalignment driven by the system committing to an interpretation that has not been adequately clarified.

    From the RVT perspective, the correct response to a Rogue Variable is not necessarily to increase confidence via sharper point estimates, but to represent and preserve the structured ambiguity until interaction (often human-in-the-loop) resolves it. This motivates a modeling framework in which ambiguity is representable as a coherent configuration, and in which \emph{collapse} corresponds to an informational decoherence event driven by clarification rather than an automatic maximization step.

    \subsection{Motivation for a quantum-consistent representation}
    
    The limitations above suggest that the pre-event regime requires a formalism in which (i) competing tendencies can co-exist without being forced into exclusive hypotheses, (ii) inference can be context- and order-sensitive, and (iii) the transition from ambiguity to commitment can be modeled as an update induced by interaction. In the next section we introduce such a representation by embedding the Mirrored Personal Graph into a quantum-consistent state space and defining the Quantum MPG State, which provides the foundation for Hamiltonian dynamics and for operator-based Rogue Variable detection.


\section{Decoherence, collapse, and human-in-the-loop alignment}
\label{sec:decoherence}

    \subsection{Informational decoherence, not physical collapse}
    In Rogue Variable Theory (RVT), the term \emph{collapse} is used in an informational and operational sense rather than a physical one. A Rogue Variable corresponds to a pre-event configuration in which multiple interpretations or action-tendencies co-exist without a single stabilized narrative. \emph{Collapse} refers to the moment when interaction supplies enough constraint to resolve this ambiguity into a more determinate representation.

    \medskip

    Concretely, collapse is an \emph{information update under interaction}. In this sense, it is closer to interpretational stances in quantum information (e.g. QBism- and relational perspectives) than to claims about physical wavefunction collapse in biological substrates \cite{FuchsMerminSchack2014,Rovelli1996}. The formalism is used because it naturally represents contextuality and order sensitivity; it does not require any assumption of quantum coherence in neural tissue \cite{Tegmark2000}.

    \medskip

    Within this interpretation, \emph{measurement} is any interaction that reduces the space of compatible meanings: a question, a reframing, a commitment to a value, or an explicit acknowledgment of affect. Context plays a constitutive role: the same pre-event state can resolve differently depending on how it is probed.

    \subsection{Human-in-the-loop as an alignment mechanism}
    In many high-stakes settings, the resolution of meaning is not something an AI system should perform autonomously. Humans resolve ambiguity by answering questions, reframing context, acknowledging emotion, and choosing values. RVT treats this not as an external exception but as an essential component of cognition and alignment.

    \medskip

    From an engineering viewpoint, this provides a principled interface for human-in-the-loop operation: when the system detects a pre-event regime (high ambiguity, unstable inference, or elevated divergence), it should preferentially shift from \emph{answering} to \emph{clarification}, eliciting the contextual and normative information required for a responsible update. In this way, human interaction functions as an alignment mechanism that reduces premature commitment, mitigates overconfidence, and helps ensure that subsequent actions reflect the user's intended meaning and values \cite{Amershi2019,Holzinger2016}.

    \subsection{Connection to the mathematical model}
    The next sections introduce a quantum-consistent state-space representation in which pre-event configurations are represented explicitly as states, and where interaction-driven updates can be modeled as context-dependent transformations of those states. In particular, the distinction between smooth evolution (between interactions) and clarification-driven updates (during interaction) motivates the separation between Hamiltonian dynamics and the error- or divergence-driven operators used later for Rogue Variable detection.


\section{Relationship to quantum physics: scope and non-claims}
\label{sec:quantum_physics_relationship}

    \subsection{What RVT does not claim}
    Rogue Variable Theory (RVT) adopts quantum-consistent mathematical structure as a modeling language for contextual inference. It does \emph{not} make claims about the physical substrate of cognition. In particular, RVT does not claim that:
        \begin{itemize}
            \item the brain is a quantum computer,
            \item neurons maintain long-lived quantum coherence,
            \item consciousness collapses physical wavefunctions,
            \item microtubules or other specific quantum-biological mechanisms are required for cognition \cite{HameroffPenrose2014,Fisher2015}.
        \end{itemize}
    Such hypotheses are speculative and are not assumed by the framework developed here. The present work is compatible with fully classical implementations and should be understood as an information-theoretic and operational modeling proposal.

    \subsection{What RVT is consistent with}
    While RVT is not quantum-fundamental, it is consistent with a number of well-established ideas from quantum probability and quantum information theory that are relevant as \emph{formal} tools for modeling contextual cognition. In particular, RVT is consistent with:
        \begin{itemize}
            \item \textbf{Quantum probability} as a non-classical probability calculus suitable for contextual systems,
            \item \textbf{Contextuality} understood as an informational property of inference under incompatible measurement contexts,
            \item \textbf{Non-commutative updates} in which the order of queries or interactions can change the resulting state,
            \item \textbf{Open-system analogies} in which interactions induce effective decoherence and stabilize particular outcomes,
            \item \textbf{Epistemic interpretations} of quantum information, in which the formalism encodes knowledge, constraints, and interaction structure rather than microscopic ontology.
        \end{itemize}

    In this sense, RVT is \emph{quantum-consistent}: it uses Hilbert-space geometry, operators, and update rules to represent ambiguity, context dependence, and interaction-driven resolution \cite{BusemeyerBruza2012,PothosBusemeyer2013,Khrennikov2010,AbramskyBrandenburger2011,DzhafarovKujala2016,NielsenChuang2010,Zurek2003,Schlosshauer2004}. The framework remains agnostic about whether any physical quantum processes play a role in cognition and does not require such processes to be true in order to be valid and useful.


\section{Could RVT ever become physically quantum?}
\label{sec:RVT_vs_QP}

    \subsection{The only defensible path}
    Although RVT is presented here as a quantum-consistent information-theoretic framework, one may ask whether it could ever connect to claims about physical quantum processes in biological cognition. Any such connection would require an explicit and testable empirical program. In particular, a defensible path would require:
        \begin{itemize}
            \item \textbf{Specific biological substrates} to be hypothesized (with a clear account of the relevant degrees of freedom and interaction mechanisms),
            \item \textbf{Measurable coherence-like timescales} or other operational signatures that can be experimentally probed in the proposed substrates,
            \item \textbf{Ruling out classical explanations} (including classical stochastic and dynamical models) for the same behavioral and physiological phenomena,
            \item \textbf{Distinct predictions} that differ from what classical models would produce, with experimental designs capable of discriminating between them.
        \end{itemize}
    Establishing such claims would require a separate research program in quantum biology and cognitive neuroscience, with explicit measurement protocols and falsifiable predictions. It is not assumed in the present work.

    \subsection{Why physical quantumness is not required}
    Crucially, RVT does not depend on physical quantum processes in order to be valid, useful, or true as a modeling framework. The value of RVT lies in the \emph{information structure} required to represent pre-event cognition: contextuality, order dependence, and the interaction-driven resolution of structured ambiguity. These can be modeled rigorously using quantum-consistent mathematics while remaining fully compatible with classical computation and classical biological substrates.


\section{Quantum MPG representation and graph Hilbert space}
\label{sec:math:Hilbert}
    
    \subsection{Mirrored Personal Graph and metrics}
    We model a user's evolving cognitive configuration by a time-indexed \emph{Mirrored Personal Graph} ($\MPG$) with onboarding time $t_0\in\R$. For each $t\in\R$, let
        \begin{equation}
            \label{eq:MPG-snapshot}
            G_t=(V_t,E_t),
        \end{equation}
    where $V_t$ is a finite set of nodes active at time $t$ and $E_t\subseteq V_t\times V_t$ is a finite set of directed edges. The $\MPG$ may be multi-level: segments at one level may be lifted into nodes at a higher level. Depending on the application, the system may work with a fixed level or with a union of multiple levels.

    Each node and edge is equipped with metrics represented numerically as feature vectors
        \begin{equation}
            \label{eq:node-edge-metrics}
            m_t(v)\in\R^{d_v},\quad v\in V_t,\qquad m_t(e)\in\R^{d_e},\quad e\in E_t.
        \end{equation}
    These metrics may encode interpretable attributes such as importance, confidence, valence, recency, stability, edge type, layer labels, or embedded semantic descriptors \cite{Shuman2013,SandryhailaMoura2013}.

    \subsection{Graph Hilbert space}
    To place time-varying graphs into a single state space, we assign a global index to each $\MPG$ node through an enumerated universe
        \begin{equation*}
            V = \{v_1, v_2, v_3, \dots\}.
        \end{equation*}
    Whenever a new node is created, it is assigned the next unused label $v_i \in V$, and this label is not reused for any other node. At any time $t$, the set of active nodes $V_t$ can be identified with a finite subset of $V$, and the corresponding edge set satisfies $E_t \subseteq V_t \times V_t$.

    The graph Hilbert space is then defined as the $\ell^2$-space over the (global) node set $V$:
        \begin{equation*}
            \mathcal{H}_{\mathrm{graph}} := \ell^2(V)
            = \Bigl\{\psi : V \to \mathbb{C} \;\Big|\; \sum_{v_i \in V} |\psi(v_i)|^2 < \infty \Bigr\}.
        \end{equation*}
    This is an infinite-dimensional separable Hilbert space, and the inner product on $\mathcal{H}_{\mathrm{graph}}$ is defined by
        \begin{equation}
            \langle \phi \mid \psi \rangle = \sum_{v_i \in V} \phi(v_i)^* \,\psi(v_i), \qquad \phi, \psi \in \mathcal{H}_{\mathrm{graph}}.
        \end{equation}
    We denote by $\{ \ket{v_i} : v_i \in V \}$ the canonical orthonormal basis of $\mathcal{H}_{\mathrm{graph}}$, where
        \begin{equation}
            \langle v_i \mid v_j \rangle = \delta_{ij}, \qquad \psi(v_i) = \langle v_i \mid \psi\rangle \quad \text{for all } \psi \in \mathcal{H}_{\mathrm{graph}}.
        \end{equation}

    For a given time $t$, the instantaneous Mirrored Personal Graph $G_t = (V_t, E_t)$ is treated as a finite subgraph of the global node universe $V$, and the corresponding state $\ket{\Psi_t} \in \mathcal{H}_{\mathrm{graph}}$ has support only on nodes in $V_t$, i.e.
        \begin{equation*}
            \ket{\Psi_t} = \sum_{v_i \in V_t} \psi_t(v_i)\,\ket{v_i}, \qquad \psi_t(v_i) = 0 \text{ for all } v_i \notin V_t.
        \end{equation*}
    Each node $v_i$ is associated with a time horizon $[t_{\mathrm{start}}(v_i), \infty)$, which is one of the node metrics, and for times $t$ outside this interval the corresponding amplitude is set to zero, $\psi_t(v_i) = 0$ for $t < t_{\mathrm{start}}(v_i)$. This construction embeds all time-varying MPG instances $\{G_t\}$ into a single, fixed Hilbert space with a fixed orthonormal basis, while allowing different subsets of basis vectors to be active at different times.
    
    \subsection{Quantum MPG State (QMS)}
    The system represents the user's instantaneous configuration over the Mirrored Personal Graph by a normalized vector $\ket{\Psi_t} \in \mathcal{H}_{\mathrm{graph}}$, referred to as the \emph{Quantum MPG State} (QMS) at time $t$. The QMS is constructed from node metrics $m_t(v_i)$, which may include intrinsic properties (such as importance, confidence, valence, recency, stability) and topological or geometric properties of the MPG (such as degree, centrality scores, path counts or alignment measures with respect to other nodes and segments).

    For each active node $v_i \in V_t$, the system computes a complex amplitude
        \begin{equation}
            \label{eq:Psi-map}
            \psi_t(v_i) = \Psi_{\mathrm{map}}\bigl(m_t(v_i), \mathcal{C}_t\bigr),
        \end{equation}
    where $\Psi_{\mathrm{map}}$ is a mapping from node metrics (and optionally a contextual configuration $\mathcal{C}_t$) to amplitudes, and sets $\psi_t(v_i) = 0$ for all $v_i \notin V_t$. The Quantum MPG State is then given by
        \begin{equation}
            \label{eq:QMS}
            \ket{\Psi_t} := \sum_{v_i \in V} \psi_t(v_i)\,\ket{v_i},
        \end{equation}
    and the mapping $\Psi_{\mathrm{map}}$ is chosen such that $\ket{\Psi_t}$ is normalized, i.e.\ $\braket{\Psi_t|\Psi_t} = 1$.

    \begin{itemize}
        \item Each node $v_i \in V$ corresponds to a cognitive entity (for example, a belief, routine, somatic pattern, identity fragment, memory or lifted segment), and the amplitude $\psi_t(v_i)$ reflects how strongly that entity is inferred to be active, salient or contextually relevant under a given QMS at time $t$. The squared magnitude $|\psi_t(v_i)|^2$ is interpreted as a normalized activation weight assigned to $v_i$, so nodes with larger values are those the system currently considers more central in explaining or predicting the user's state or behavior in the chosen context.
        
        \item The metrics $m_t(v_i)$ are updated over time based on multimodal sensor data and internal model signals. Let $x_t$ denote the collection of exogenous and endogenous inputs at time $t$, for example,
            \begin{equation*}
                x_t :=  \bigl(x_t^{\mathrm{phys}}, x_t^{\mathrm{beh}}, x_t^{\mathrm{ctx}}, x_t^{\mathrm{model}}\bigr),
            \end{equation*}
        where $x_t^{\mathrm{phys}}$ may include physiological measurements from wearable sensors (e.g.\ heart rate, sleep or activity signals), $x_t^{\mathrm{beh}}$ may include behavioral and interaction logs, $x_t^{\mathrm{ctx}}$ may include contextual information (time of day, location, application usage), and $x_t^{\mathrm{model}}$ may include internal model outputs (e.g.\ prediction errors, uncertainty estimates or latent embeddings). The system combines $x_t$ with historical MPG information and updates the metrics for each node. As these metrics change, the amplitudes $\psi_t(v_i)$ and thus the Quantum MPG State $\ket{\Psi_t}$ are updated accordingly. 

        \item The system may instantiate multiple QMS variants by choosing different mappings $\Psi_{\mathrm{map}}$ or different contextual configurations $\mathcal{C}_t$. For example, one QMS may emphasize metrics and subgraphs related to different aspects of cognition or activity of the user. In each case, the same underlying MPG and node metrics can be projected into different quantum-inspired views by reweighting or filtering the metrics used in $\Psi_{\mathrm{map}}$, allowing the system to compare and analyze several context-specific Quantum MPG States at the same time or over the same time interval.
    \end{itemize}

    The QMS therefore provides a compact, Hilbert-space representation of the user's inferred MPG-based state at time $t$, which can be specialized to different aspects of the user's life or behavior by appropriate choices of $\Psi_{\mathrm{map}}$ and $\mathcal{C}_t$. Between such updates, $\ket{\Psi_t}$ may evolve according to a Hamiltonian constructed from edge-level information, as described in the following section.


\section{Hamiltonian dynamics on the graph Hilbert space}
\label{sec:math:Hamiltonian}

    \subsection{Schr\"odinger-type evolution and prediction}
    Between observation-driven reconstructions of the QMS, we model the evolution of $\ket{\Psi_t}$ through a time-dependent Hamiltonian operator
        \begin{equation}
            \label{eq:H-selfadjoint}
            \hat{H}_t:\mathcal{H}_{\mathrm{graph}}\to\mathcal{H}_{\mathrm{graph}}, \qquad \hat{H}_t^\dagger=\hat{H}_t.
        \end{equation}
    Self-adjointness implies that the induced time-evolution operators are unitary and preserve the norm of the quantum-inspired state. The evolution of the QMS is governed by Schr\"odinger-type dynamics. For a small increment $\Delta t>0$, treating $\hat{H}_t$ as locally constant on $[t,t+\Delta t]$, define
        \begin{equation}
            \label{eq:unitary-step}
            U_{t,t+\Delta t}:=\exp\!\bigl(-i\,\hat{H}_t\,\Delta t\bigr), \qquad \ket{\Psi_{t+\Delta t}}=U_{t,t+\Delta t}\ket{\Psi_t}.
        \end{equation}
    In differential form (see \cite{FarhiGutmann1998,Childs2009,Aharonov1993}),
        \begin{equation}
            \label{eq:schrodinger}
            i\,\frac{d}{dt}\ket{\Psi_t}=\hat{H}_t\ket{\Psi_t}.
        \end{equation}
    When new observations arrive, the system updates metrics $m_t(\cdot)$ and reconstructs an updated state via \eqref{eq:Psi-map}--\eqref{eq:QMS}. The mismatch between the Hamiltonian prior \eqref{eq:unitary-step} and the observation-updated QMS yields a natural divergence or prediction-error signal used later for Rogue Variable detection.

    \subsection{Hamiltonian construction from edge-level couplings}
    Edge information is aggregated into directed edge-metric vectors
        \begin{equation}
            \label{eq:edge-metrics-directed}
            m_t^{\rightarrow}(v_i,v_j)\in\R^{d_e},\qquad (v_i,v_j)\in V\times V.
        \end{equation}
    From these, the system defines a complex coupling kernel $J_t:V\times V\to\C$ constrained by Hermitian symmetry
        \begin{equation}
            \label{eq:J-hermitian}
            J_t(v_i,v_j)=\overline{J_t(v_j,v_i)}.
        \end{equation}
    In addition, node-local terms $h_t:V\to\R$ are constructed from node metrics.

    A canonical Hamiltonian operator is then
        \begin{equation}
            \label{eq:H-global}
            \hat{H}_t := \sum_{v_i,v_j\in V}J_t(v_i,v_j)\ket{v_i}\bra{v_j} + \sum_{v_i\in V}h_t(v_i)\ket{v_i}\bra{v_i}.
        \end{equation}
    In practice $J_t(v_i,v_j)$ is sparse and typically nonzero primarily when $v_i,v_j\in V_t$ and the $\MPG$ contains relevant edges between them (in either direction). Under \eqref{eq:J-hermitian} and $h_t(v_i)\in\R$, the operator $\hat{H}_t$ is self-adjoint.

    \subsection{Baseline eigenmodes as latent dynamical factors}
    In many applications it is useful to construct a baseline or reference Hamiltonian $\hat{H}_{\mathrm{base}}$ from a long-run aggregated $\MPG$ representing typical dynamics. Its spectral decomposition defines a latent factor basis (\cite{Chung1997}). Consider eigenpairs
        \begin{equation}
            \label{eq:baseline-eigs}
            \hat{H}_{\mathrm{base}}\ket{\phi_k}=E_k\ket{\phi_k},\qquad E_k\in\R,
        \end{equation}
    with $\{\ket{\phi_k}\}$ orthonormal (in the finite-dimensional truncation) or an orthonormal set (in the infinite-dimensional setting). Any normalized QMS admits an expansion
        \begin{equation}
            \label{eq:latent-expansion}
            \ket{\Psi_t}=\sum_k c_k(t)\ket{\phi_k}, \qquad c_k(t):=\braket{\phi_k\mid\Psi_t}, \qquad \sum_k|c_k(t)|^2=1.
        \end{equation}
    The coefficients $|c_k(t)|^2$ quantify activation of baseline latent factors and provide an interpretable coordinate system for tracking dynamical drift and for defining cross-user reference mappings in the Rosetta Stone Layer.


\section{Rogue Variables in the Hilbert-space model}
\label{sec:math:RV}

    This section provides an operational definition of Rogue Variables in the previously introduced QMS/Hamiltonian architecture. The guiding principle is: Rogue Variables are those structured components of the QMS trajectory that are disproportionately associated with divergence from baseline predictions or with downstream error.

    \subsection{Error signal and time-indexed divergence}
    Let $\{\ket{\Psi_t}:t\in\mathcal{T}\}$ be a collection of QMS snapshots on a time index set $\mathcal{T}$ (finite or discretized). Let $\epsilon_t\in\R_{\ge 0}$ denote a scalar error or divergence signal at time $t$. Depending on the application, $\epsilon_t$ may represent: behavioral prediction error, loss of coherence with respect to a baseline manifold, mismatch between the Hamiltonian-predicted prior and the observation-updated QMS, or a composite risk score combining such components.

    \subsection{The error-weighted rogue operator}
    We define an error-weighted operator on $\mathcal{H}_{\mathrm{graph}}$ by
        \begin{equation}
            \label{eq:O-epsilon}
            \hat{O}_\epsilon :=\frac{1}{Z_\epsilon}\sum_{t\in\mathcal{T}}\epsilon_t\,\ket{\Psi_t}\bra{\Psi_t}, \qquad Z_\epsilon:=\sum_{t\in\mathcal{T}}\epsilon_t>0.   
        \end{equation}  
    The operator $\hat{O}_\epsilon$ is positive semi-definite and self-adjoint. Intuitively, it acts as a covariance-like object that emphasizes state-space directions frequently visited during high-error periods.

    \subsection{Rogue factor directions}
    Let $(\mu_j,\ket{\chi_j})$ be eigenpairs of $\hat{O}_\epsilon$:
        \begin{equation}
            \label{eq:rogue-eigs}
            \hat{O}_\epsilon\ket{\chi_j}=\mu_j\ket{\chi_j}, \qquad \mu_j\ge 0,
        \end{equation}
    with eigenvalues ordered non-increasingly. Eigenvectors with large $\mu_j$ represent directions along which QMS trajectories align when error is high. We refer to these eigenvectors as \emph{rogue factor directions}.

    \medskip

    When a baseline eigenbasis $\{\ket{\phi_k}\}$ from \eqref{eq:baseline-eigs} is available, we may express rogue directions in the latent factor basis:
        \begin{equation}
            \label{eq:rogue-in-latent}
            \ket{\chi_j}=\sum_k \alpha_{jk}\ket{\phi_k}, \qquad \alpha_{jk}:=\braket{\phi_k\mid\chi_j}.
        \end{equation}
    This decomposition quantifies how divergence distributes across baseline dynamical modes.

    \subsection{Candidate Rogue Variable segments}
    Expanding a rogue direction in the node basis yields
        \begin{equation}
            \label{eq:rogue-in-node}
            \ket{\chi_j}=\sum_{v_i\in V}\beta_j(v_i)\ket{v_i}, \qquad \beta_j(v_i):=\braket{v_i\mid\chi_j}.
        \end{equation}
    The coefficients $\beta_j(v_i)$ indicate how strongly each node participates in rogue direction $j$. For a loading threshold $\theta_{\mathrm{load}}>0$, define the high-loading node set
        \begin{equation}
            \label{eq:rogue-node-set}
            S_j:=\Bigl\{v_i\in V:\ |\beta_j(v_i)|\ge \theta_{\mathrm{load}}\Bigr\}.
        \end{equation}
    In practice, $S_j$ may be refined by restricting to connected subgraphs, existing $\MPG$ segments, or semantically coherent subsets according to node metrics.

    \medskip
    
    Each candidate segment $S\subseteq V$ is represented at the operator level by the orthogonal projector
        \begin{equation}
            \label{eq:proj-S}
            P_S:=\sum_{v_i\in S}\ket{v_i}\bra{v_i}.
        \end{equation}

    \subsection{Ablation criterion and operational definition}
    To evaluate whether a candidate segment $S$ behaves as a Rogue Variable, we ablate its contribution from the QMS. For times $t$ such that $(I-P_S)\ket{\Psi_t}\neq 0$, define the ablated state
        \begin{equation}
            \label{eq:ablated-state}    
            \ket{\Psi_t^{(\neg S)}} :=\frac{(I-P_S)\ket{\Psi_t}}{\|(I-P_S)\ket{\Psi_t}\|},
        \end{equation}
    where $I$ denotes the identity on $\mathcal{H}_{\mathrm{graph}}$. Recompute the relevant error/divergence measure using $\ket{\Psi_t^{(\neg S)}}$ in place of $\ket{\Psi_t}$, yielding a modified signal $\epsilon_t^{(\neg S)}$. Define the average error reduction
        \begin{equation}
            \label{eq:Delta-S}
            \Delta(S) :=\frac{1}{|\mathcal{T}|}\sum_{t\in\mathcal{T}}\bigl(\epsilon_t-\epsilon_t^{(\neg S)}\bigr).
        \end{equation}
    A positive value of $\Delta(S)$ indicates that removing the influence of $S$ improves performance or reduces divergence.

    \begin{definition}[Rogue Variable segment (operational)]
        A candidate segment $S\subseteq V$ is called a \emph{Rogue Variable segment} on the time window $\mathcal{T}$ if it is supported by a rogue direction (e.g.\ $S$ is derived from high-loadings in \eqref{eq:rogue-node-set} for some leading eigenvector $\ket{\chi_j}$) and its ablation yields a positive reduction score $\Delta(S)>0$ as in \eqref{eq:Delta-S}.
    \end{definition}

    This operator-based criterion links the conceptual notion of structured pre-event ambiguity to a computable mechanism: rogue operators identify divergence-associated directions, and their high-loading nodes identify segments whose removal reduces error. These segments are then candidates for interpretability-driven inspection and for human-in-the-loop clarification interventions.


\section{Rosetta Stone Layer alignment algorithm}
\label{sec:math:Rosetta}

    The Rosetta Stone Layer (RSL) provides a mechanism for cross-user comparison and aggregation without requiring explicit node-level alignment between personal graphs. The core idea is to use user-specific Hamiltonian eigenstructures as a coordinate interface and to map these coordinates into a shared reference space.

    \subsection{From individuals to collectives}
    Rogue Variables are defined at the level of an individual trajectory: they arise as structured, pre-event configurations within a person's evolving cognitive state. However, many of the most important transitions occur not only within individuals but also within groups. When multiple individuals interact---in teams, organizations, communities, or societies-ambiguity can become \emph{shared}, and transitions can become \emph{synchronized}. In such settings, one can observe collective phenomena such as coordinated shifts in interpretation, cascading commitment events, and the emergence of group-level attractors.

    \medskip

    To study such effects, we require a mechanism for aggregating and comparing pre-event configurations across individuals without assuming that personal representations share a common node vocabulary. The Rosetta Stone Layer (RSL) is introduced precisely for this purpose: it constructs a collective cognitive manifold by mapping user-specific latent structures into a shared reference space. In this shared geometry, individual trajectories can be analyzed together, enabling detection of collective Rogue Variables and group-level transition structure.

    \subsection{Universal cognitive signatures} 
    Once representations from multiple users are mapped into a common reference geometry, the system can search for \emph{universal cognitive signatures} that recur across populations. Examples include:
        \begin{itemize}
            \item recurring pre-event patterns that precede decisions, conflicts, or commitments,
            \item common ``collapse paths'' by which ambiguity resolves under interaction,
            \item archetypal configurations of ethical tension (e.g.\ value conflicts that require clarification),
            \item structured signatures of creative emergence in which competing interpretations reorganize into new stable forms,
            \item early indicators of group instability, such as synchronized divergence and correlated ambiguity across individuals.
        \end{itemize}
    In this view, the RSL supports the development of a \emph{grammar of cognition}: a library of pre-event motifs and transition patterns that can be recognized, compared, and used for forecasting and intervention. Importantly, these signatures are defined at the level of state-space geometry and operator structure rather than at the level of literal node identity, which is typically idiosyncratic across individuals.

    \subsection{User-specific latent factor representations}    
    For each user $u\in\mathcal{U}$, let $\mathcal{H}_{\mathrm{graph}}^{(u)}$ be the user's graph Hilbert space and let
        \begin{equation}
            \label{eq:user-H}
            \hat{H}^{(u)} : \mathcal{H}^{(u)}_{\mathrm{graph}} \to\allowbreak \mathcal{H}^{(u)}_{\mathrm{graph}}
         \end{equation}
    be a chosen self-adjoint operator (typically a baseline Hamiltonian). Its eigen-decomposition becomes
        \begin{equation}
            \label{eq:user-eigs}
            \hat{H}^{(u)}\ket{\phi_k^{(u)}}=E_k^{(u)}\ket{\phi_k^{(u)}},\qquad k\in\N,
        \end{equation}
    with $\{\ket{\phi_k^{(u)}}\}$ orthonormal. In implementation one works with a finite truncation $K^{(u)}\subset\N$ containing the dominant modes.

    \medskip

    Any state $\ket{\Psi^{(u)}}\in\mathcal{H}_{\mathrm{graph}}^{(u)}$ admits an expansion
        \begin{equation}
            \label{eq:user-expansion}
            \ket{\Psi^{(u)}}=\sum_{k\in\N} c_k^{(u)}(\Psi)\ket{\phi_k^{(u)}}, \qquad c_k^{(u)}(\Psi):=\braket{\phi_k^{(u)}\mid\Psi^{(u)}}.
        \end{equation}
    In particular, the Quantum MPG States (QMS) $\ket{\Psi_t^{(u)}}$, rogue directions $\ket{\chi_j^{(u)}}$, and operators such as Hamiltonians and Rogue Variable projectors may all be represented in this eigenbasis.

    \subsection{Rosetta Cognitive Reference Space, coefficient extraction and alignment transforms}
    Define a shared reference index set $W=\{w_1,w_2,\dots\}$ and the reference Hilbert space
        \begin{equation}
            \label{eq:H-ref}
            \mathcal{H}_{\mathrm{ref}}:=\ell^2(W),
        \end{equation}
    with canonical basis $\{\ket{e_i}:i\in\N\}$. Thus any $\ket{\Phi}\in\mathcal{H}_{\mathrm{ref}}$ can be written as
        \begin{equation}
            \label{eq:ref-vector}   
            \ket{\Phi}=\sum_{i\in\N}\Phi_i\,\ket{e_i},\qquad \sum_{i\in\N}|\Phi_i|^2<\infty.
        \end{equation}
    For each user $u$, define a coefficient extraction map
        \begin{equation}
            \label{eq:C-u}
            C^{(u)}:\mathcal{H}_{\mathrm{graph}}^{(u)}\to\mathcal{H}_{\mathrm{ref}}
        \end{equation}
    that places user-specific eigen-coefficients into reference coordinates. One simple realization uses an injective index mapping $\iota^{(u)}:\N\to\N$ and sets
        \begin{equation}
            \label{eq:C-u-definition}
            C^{(u)}\ket{\Psi^{(u)}} :=\sum_{k\in\N} c_k^{(u)}(\Psi)\,\ket{e_{\iota^{(u)}(k)}}.
        \end{equation}
    In practice, only $k\in K^{(u)}$ are used, so the result is sparse in the reference basis.

    \medskip
    
    To account for the fact that user eigenmodes need not be semantically aligned across users, introduce a user-specific alignment transform
        \begin{equation}
            \label{eq:R-u}
            R^{(u)}:\mathcal{H}_{\mathrm{ref}}\to\mathcal{H}_{\mathrm{ref}},
        \end{equation}
    typically unitary or approximately orthogonal. The overall RSL alignment operator is
        \begin{equation}
            \label{eq:A-u}
            A^{(u)}:=R^{(u)}\circ C^{(u)}:\mathcal{H}_{\mathrm{graph}}^{(u)}\to\mathcal{H}_{\mathrm{ref}}.
        \end{equation}
    Applied to a user QMS $\ket{\Psi_t^{(u)}}$, the aligned reference state is
        \begin{equation}
            \label{eq:Phi-u}
            \ket{\Phi_t^{(u)}}:=A^{(u)}\ket{\Psi_t^{(u)}}\in\mathcal{H}_{\mathrm{ref}}.
        \end{equation}

    \subsection{Mapping operators and rogue structures into the reference space}
    The same alignment operators map user-specific operators to the reference space. For a bounded operator $\hat{O}^{(u)}$ on $\mathcal{H}_{\mathrm{graph}}^{(u)}$, define
        \begin{equation}
            \label{eq:O-ref}
            \hat{O}_{\mathrm{ref}}^{(u)}:=A^{(u)}\hat{O}^{(u)}A^{(u)\dagger}.
        \end{equation}
    In particular, user Hamiltonians, rogue operators, rogue directions, and Rogue Variable projectors can be transported into $\mathcal{H}_{\mathrm{ref}}$:
        \begin{eqnarray*}
            \hat{H}_{\mathrm{ref}}^{(u)} &:=& A^{(u)}\hat{H}^{(u)}A^{(u)\dagger},\quad \hat{O}_{\epsilon,\mathrm{ref}}^{(u)}:=A^{(u)}\hat{O}_{\epsilon}^{(u)}A^{(u)\dagger},\\
            \ket{\zeta_j^{(u)}} &:=& A^{(u)}\ket{\chi_j^{(u)}},\quad P_{S,\mathrm{ref}}^{(u)}:=A^{(u)}P_S^{(u)}A^{(u)\dagger}.
        \end{eqnarray*}

    \subsection{Cross-user comparison and projection back to individuals}
    Once embedded in $\mathcal{H}_{\mathrm{ref}}$, aligned states $\ket{\Phi_t^{(u)}}$ can be compared across users using inner products or derived distances, enabling discovery of shared pre-event configurations and recurring transition patterns at a meta-level. Likewise, aligned rogue directions and mapped operators support aggregation of group-level \emph{rogue signatures}.

    \medskip

    Reference-space findings can be projected back to an individual user by adjoint maps. Given a reference direction $\ket{\zeta}\in\mathcal{H}_{\mathrm{ref}}$ or an operator $\hat{O}_{\mathrm{ref}}$ on $\mathcal{H}_{\mathrm{ref}}$, define user-specific counterparts
        \begin{equation}
            \label{eq:project-back}
            \ket{\chi^{(u)}}:=A^{(u)\dagger}\ket{\zeta}, \qquad \hat{O}^{(u)}:=A^{(u)\dagger}\hat{O}_{\mathrm{ref}}A^{(u)}.
        \end{equation}
    Expanding $\ket{\chi^{(u)}}$ in the user's node basis highlights the personal $\MPG$ nodes and segments involved, preserving interpretability and supporting personalized monitoring and human-in-the-loop clarification.

    \subsection{Choosing alignment transforms}  
    The alignment transforms $R^{(u)}$ may be optimized so that (i) states corresponding to similar situations across users become close in $\mathcal{H}_{\mathrm{ref}}$, and (ii) latent factors with analogous semantic meaning occupy similar reference directions (\cite{Schonemann1966,Kabsch1976,GowerDijksterhuis2004}). A natural signal for such optimization is the \emph{semantic fingerprint} of eigenmodes: writing
        \begin{equation*}
            \ket{\phi_k^{(u)}}=\sum_{v_i^{(u)}\in V^{(u)}}\alpha_k^{(u)}(v_i^{(u)})\ket{v_i^{(u)}},
        \end{equation*}
    the coefficients $\alpha_k^{(u)}(v_i^{(u)})$ together with node metrics $m^{(u)}(v_i^{(u)})$ can be used to align modes across users. This yields a data-driven Rosetta mapping that supports cross-user inference while remaining compatible with user-specific graph semantics.


\section{Discussion and potential applications}
\label{sec:applications}

The RVT framework is intended as a formal apparatus for representing and interrogating pre-event regimes:
time windows in which multiple latent interpretations remain active and context-sensitive updates are expected.
Operationally, these regimes may be indicated by elevated divergence signals $\epsilon_t$, concentration of mass
in the error-weighted operator $\hat{O}_\epsilon$, and the emergence of high-loading node sets in leading rogue
directions.

\subsection{Potential application patterns (non-exhaustive)}
Without committing to any specific domain, the following usage patterns are natural within the proposed
architecture:
\begin{itemize}
    \item \textbf{Clarification policies.} When $\epsilon_t$ is elevated and $\hat{O}_\epsilon$ concentrates on a
    small number of rogue directions, an interactive system can preferentially shift from producing point
    estimates to asking clarifying questions that reduce ambiguity and stabilize context.
    \item \textbf{Interpretability via candidate segments.} High-loading node sets extracted from rogue directions
    yield candidate segments $S$ that can be inspected (or summarized) and tested through the ablation criterion
    $\Delta(S)$, providing a structured mechanism for localizing which parts of a representation disproportionately
    contribute to divergence.
    \item \textbf{Cross-user comparison at the operator level.} The Rosetta Stone Layer enables comparison of
    trajectories and rogue directions in a shared reference space without requiring literal node-level alignment,
    supporting aggregation of recurring geometric/operator signatures across individuals.
\end{itemize}

\subsection{Empirical evaluation as future work}
Empirical validation requires (i) specifying $\Psi_{\mathrm{map}}$, $\mathcal{C}_t$, and $\epsilon_t$ for a chosen
application, (ii) constructing a dataset or interaction protocol that exposes pre-event regimes, and (iii)
evaluating whether rogue directions and ablation-selected segments improve predictive performance, interpretability,
or interactive outcomes relative to appropriate baselines. These tasks are left to future work.


\section{Positioning RVT in the scientific landscape}
\label{sec:RVT_vs_science}

    Rogue Variable Theory (RVT) is motivated by a concrete modeling gap-the representation of \emph{pre-event} cognitive states-and proposes a quantum-consistent formalism to address that gap. In doing so, RVT sits at the intersection of several scientific traditions:
        \begin{itemize}
            \item \textbf{Quantum cognition,} which provides non-classical probabilistic tools for contextuality, order effects, and interference-like phenomena in judgment and decision;
            \item \textbf{Predictive processing,} in which cognition is understood as inference under uncertainty with action and perception driven by prediction error and model revision;
            \item \textbf{Complex systems,} emphasizing emergent dynamics, transitions, and multi-level organization in cognitive and social processes;
            \item \textbf{Collective intelligence,} where group-level attractors and synchronized transitions arise from interacting individuals;
            \item \textbf{Human--AI symbiosis,} in which systems must treat human meaning-making and value selection as essential parts of the loop rather than as externalities.
        \end{itemize}
    Within this landscape, RVT motivates a modeling category that we refer to as \emph{pre-event cognitive modeling}: the study of structured ambiguity and transitional configurations that precede stabilized decisions, labels, and narratives. The mathematical framework developed in this paper-graph-based state representations, Hamiltonian dynamics, rogue-direction operators, and Rosetta-layer alignment-is intended as a formal starting point for this category.


\section{Conclusion}    
\label{sec:conclusion}

    This paper developed Rogue Variable Theory (RVT) as a rigorous framework for understanding and operationalizing one of the most important-and least well represented-phases of cognition: the moment \emph{before} meaning forms. Rogue Variables were introduced conceptually as structured pre-event configurations, and then instantiated mathematically in a quantum-consistent architecture: a Mirrored Personal Graph embedded into a graph Hilbert space, a normalized Quantum MPG State updated from metrics and context, Hamiltonian dynamics for prediction between interactions, an error-weighted rogue operator whose spectral structure identifies divergence-associated rogue directions and candidate Rogue Variable segments, and a Rosetta Stone Layer alignment algorithm enabling cross-user aggregation in a shared reference geometry.

    \medskip

    The framework does not mystify cognition with physics, nor does it reduce humans to statistics. Instead, it uses a representation suited to contextual inference and structured ambiguity, while preserving human agency through clarification-driven updates and human-in-the-loop resolution of meaning. RVT is not a theory of quantum physics; it is, however, compatible with quantum-informational views in which the formalism is understood as an epistemic and interaction-sensitive structure. In this sense, RVT offers an information-theoretic foundation for studying cognition under uncertainty, designing alignment-aware AI systems, and analyzing collective transitions in groups and organizations.

    \subsection*{Outlook}
    Immediate next steps include application-specific instantiation of $\Psi_{\mathrm{map}}$ and $\epsilon_t$, finite-dimensional truncation and numerical stability analysis of the operators involved, and empirical studies testing whether rogue directions and ablation-selected segments improve prediction, interpretability, or clarification efficiency in interactive settings.

\printbibliography
    
\end{document}